\documentclass[10pt,letterpaper]{article}
\usepackage{opex3}

\begin{document}

\title{\Large Spatial dispersion and energy in strong chiral medium}

\author{Chao Zhang$^{1,2}$ and Tie Jun Cui$^{1*}$}

\address{$^{1}$ Center for Computational Electromagnetics and the
State Key Laboratory of Millimeter Waves, \\ Department of Radio
Engineering, Southeast University, Nanjing 210096, P. R. China.
\\
$^{2}$Dept. of Electrical and Computer Engineering, University of
New Mexico, Albuquerque, NM 87131, USA.}

\email{czhang@unm.edu, tjcui@seu.edu.cn}


\begin{abstract}
Since the discovery of backward-wave materials, people have tried
to realize strong chiral medium, which is traditionally thought
impossible mainly for the reason of energy and spatial dispersion.
We compare the two most popular descriptions of chiral medium.
After analyzing several possible reasons for the traditional
restriction, we show that strong chirality parameter leads to
positive energy without any frequency-band limitation in the weak
spatial dispersion. Moreover, strong chirality does not result in
a strong spatial dispersion, which occurs only around the
traditional limit point. For strong spatial dispersion where
higher-order terms of spatial dispersion need to be considered,
the energy conversation is also valid. Finally, we show that
strong chirality need to be realized from the conjugated type of
spatial dispersion.
\end{abstract}




\section{{{\bf Introduction}}}

Chirality is first referred to a kind of asymmetry in geometry and
group theory in mathematics. The asymmetry exists broadly in
organic molecules, crystal lattices, and liquid crystals, leading
to two stereoisomers, dextrorotatory and laevorotatory, as a hot
research domain in stereochemistry. If the two stereoisomers
coexist in one molecule (mesomer), or equally different
steroisomers get mixed (raceme), there will be no special
characters other than the common magneto-dielectric. When we get
one pure steroisomer, however, interesting phenomena occur with an
incident linearly-polarized wave, which can be seen as a
superposition of two dual circularly-polarized waves. In case of
perpendicular incidence, the two different circularly polarized
waves have different phase velocities and their polarized planes
rotate oppositely. As a result, the output polarization direction
gets rotated, also known as optical activity or natural optical
rotation phenomenon, which was first observed by Arago in 1811.
For oblique incidence, the two different polarized waves will
split even the medium is isotropic, which was verified by Fresnel
using prism series made from dextrorotatory and laevorotatory
quartz \cite{pasteur,fresnel2,polarized,Eugene}. Moreover, in
elementary particle physics, chirality and asymmetry also play
important roles, but they are out of the range of this paper.

\section{{{\bf Two major electromagnetic models to describe a chiral medium}}}

The electromagnetic theoretical explanation of optical activity is
spatial dispersion \cite{landau,sommerfeld,lindell,serdyukov}.
Usually, under the weak spatial dispersion, we use the first order
(linear) approximation, which is written as
\begin{eqnarray}
\bar{D} \hskip-2.5mm &=& \hskip-2.5mm \epsilon_{DBF}\bar{E}+\epsilon_{DBF}\beta\nabla\times\bar{E},\\
\bar{B} \hskip-2.5mm &=& \hskip-2.5mm
\mu_{DBF}\bar{H}+\mu_{DBF}\beta\nabla\times\bar{H}.
\end{eqnarray}
Such a representation is named as Drude-Born-Fedorov (DBF)
relation for a natural result of linearly spatial dispersion.
Rotation terms are added to the basic constitutive relation,
standing for the spatial dispersion, whose coefficients
$\epsilon_{DBF}\beta$ or $\mu_{DBF}\beta$ can be either positive
or negative for two stereoisomer structures. Solving the
constitutive relation together with Maxwell's equations, we can
easily get two eigenwaves, which are left and right circularly
polarized with different wavevectors. There are also some other
representations, among which the most common one is deduced by
Pasteur and Tellegen as
\begin{eqnarray}
\bar{D} \hskip-2.5mm &=& \hskip-2.5mm \epsilon\bar{E}+(\chi+i\kappa)\bar{H},\\
\bar{B} \hskip-2.5mm &=& \hskip-2.5mm
\mu\bar{H}+(\chi-i\kappa)\bar{E},
\end{eqnarray}
in which electromagnetic coupling terms are added to the basic
terms. Bi-isotropy or bi-anisotropy is used for calling such
constitutive equations, according to the parameters to be scalars
or tensors. If $\kappa=0$ and $\chi\not=0$, it is the Tellegen
medium; if $\chi=0$ and $\kappa\not=0$, as the requirement of
reciprocity, it is the Pasteur medium:
\begin{eqnarray}\label{cons1}
\bar{D} \hskip-2.5mm &=& \hskip-2.5mm \epsilon\bar{E}+i\kappa\bar{H},\\
\bar{B} \hskip-2.5mm &=& \hskip-2.5mm
\mu\bar{H}-i\kappa\bar{E}.\label{cons2}
\end{eqnarray}
We pay more attention to such a chiral medium. Positive and
negative $\kappa$ values differentiate two conjugated stereoisomer
structures. We assume $\kappa>0$ in the following analysis.

Actually, the constitutive relations above are essentially
equivalent, with corresponding parameters to be \cite{lindell}
\begin{eqnarray}\label{trans1}
\epsilon_{DBF} \hskip-2.5mm &=& \hskip-2.5mm \epsilon\left(1-\frac{\kappa^2}{\mu\epsilon}\right),\\
\mu_{DBF} \hskip-2.5mm &=& \hskip-2.5mm \mu\left(1-\frac{\kappa^2}{\mu\epsilon}\right),\\
\beta \hskip-2.5mm &=& \hskip-2.5mm
\frac{\kappa}{\omega(\mu\epsilon-\kappa^2)}.\label{trans2}
\end{eqnarray}
It is clear that the parameters are different in such two
representations. Then a question may rise up: which are the
``true'' material permittivity and permeability? The answer is,
both. The concepts of permittivity and permeability are effective
coefficients derived from a mathematical model. We actually have
different mathematical models describing the same physical
material. Thus there are different effective parameters describing
the proportion of $\bar{D}$ to $\bar{E}$ and $\bar{B}$ to
$\bar{H}$. The rotation terms in DBF model include both real and
imaginary parts, resulting in a change in the real part and
creating the imaginary chiral terms in the Pasteur model, vice
versa. In other words, the difference in representations of
coupling terms lead to different permittivity and permeability
formulations.

It should be noticed that Faraday gyratory medium can also lead to
optical rotation within the plasma or ferrite under an additional
DC magnetic field \cite{ishimaru,kong}. Hence it is not natural,
and is usually referred as ``gyratory'', ``Faraday optical
rotation'', ``magneto-optical effect'', etc. However, sometimes
people do not differentiate ``chiral'' and ``gyratory''. We need
pay attention that such two types of optical rotation have
different essence and different characters \cite{kong}. Only
natural optical activity is discussed here.

\section{\bf Energy and spatial dispersion in strong chiral medium}

There is a long dispute on strong chiral medium since it was
introduced theoretically \cite{nihility}. Traditional
electromagnetic conclusions have limited us to understand strong
chirality, i.e. $\kappa^2>\mu\epsilon$ \cite{lindell,nihility},
until we see the fact that artificial Veselago's medium
\cite{vesalago} was successfully realized in certain frequency
bands \cite{shelby}. Hence, we have to ask the following question:
can strong chiral medium exist?

In Ref. [\cite{serdyukov,nihility}], the reason for traditional
restriction of chirality parameters was concluded as: 1) The
wavevector of one eigenwave will be negative; 2) The requirement
of a positive definite matrix to keep positive energy:
\begin{equation}\label{matrix}
\left[\begin{array}{cc}
\epsilon&i\kappa\\
-i\kappa&\mu
\end{array}\right].
\end{equation}
With the exploration of backward-wave medium, we know that
negative wavevector, or opposite phase and group velocities, are
actually realizable. And there is an unfortunate mathematical
error in the second reason: in linear algebra, only if it is real
and symmetric, positive definite matrix is equivalent to that all
eigenvalues should be positive. The matrix (\ref{matrix}) is a
complex one, making the analysis on restriction of positive energy
meaningless.

Actually, in a strong bi-isotropic medium with constitutive
relations as Eqs. (\ref{cons1}) and (\ref{cons2}), the energy can
be drawn as
\begin{eqnarray}
w \hskip-2.5mm &=& \hskip-2.5mm w_e+w_m\nonumber\\
 \hskip-2.5mm &=& \hskip-2.5mm \bar{D}\cdot\bar{E}/2+\bar{B}\cdot\bar{H}/2\nonumber\\
 \hskip-2.5mm &=& \hskip-2.5mm \epsilon|\bar{E}|^2/2+i\kappa\bar{H}\cdot\bar{E}+\mu|\bar{H}|^2/2-i\kappa\bar{E}\cdot\bar{H}\nonumber\\
 \hskip-2.5mm &=& \hskip-2.5mm \epsilon|\bar{E}|^2/2+\mu|\bar{H}|^2/2.
\end{eqnarray}
Even if the strong bi-isotropic medium is not frequency
dispersive, i.e. $\kappa^2>\mu\epsilon$ for whole frequency range,
the energy will still keep positive as long as the permittivity
and permeability are positive, under the weak spatial dispersion
condition. This is quite different from the Veselago's medium
since there is no bandwidth limitation and the frequency
dispersive resonances are no longer required. In another word, the
strong chiral medium does not contradict the energy conservation,
at least in the weak spatial dispersion model.

Therefore, the real reason for traditional strong-chirality
limitation is neither negative wavevector nor energy conversation.
Next we will point out two other important reasons.

First, with the assumption that $\epsilon>0$, $\mu>0$, $\kappa>0$
and $\kappa>\sqrt{\mu\epsilon}$, we easily show that
$\epsilon_{DBF}$, $\mu_{DBF}$ and $\beta$ turn to negative from
the transformation between Pasteur constitutive relations and DBF
relations shown in Eqs. (\ref{trans1})-(\ref{trans2}). This is
absolutely unacceptable before people realizing Veselago's medium.
Actually, strong chiral medium can be equivalent to Veselago's
medium for the right circularly polarized wave
\cite{nihility,sailing}. The negative $\epsilon_{DBF}$ and
$\mu_{DBF}$ have shown such a point. Hence the negative sign in
the DBF model is not strange at all, since we realize effective
double-negative with strong chirality parameter instead of
simultaneously frequency resonances. For a limiting case, the
chiral nihility \cite{nihility}, in which $\epsilon\to0$ and
$\mu\to0$ while $\kappa\not=0$, the parameters in DBF
representation become $\epsilon_{DBF}\to\infty$,
$\mu_{DBF}\to\infty$ and $\beta=-1/(\omega\kappa)$, remaining a
finite value after a simple mathematical analysis. There is no
evidence that strong chirality cannot exist in this aspect.

Second, it is the effectiveness of linear models. Similar to the
case that linear optical and electromagnetic models can no longer
deal with very strong optical intensity and electromagnetic field,
we introduce nonlinear optics to take into account the higher
order terms of polarization. If the spatial dispersion is strong
enough, the higher order coupling terms cannot be neglected as
before \cite{lindell}. People used to mistake strong chirality
with strong spatial dispersion, hence adding a limitation to
chirality parameter, $\kappa<\sqrt{\mu\epsilon}$. We believe that
this is the most probable reason. However, the strong spatial
dispersion is embodied in the DBF model, e.g. the value of
$\beta$, while the strong chirality is represented by the Pasteur
model, e.g. the ratio of $\kappa$ to $\sqrt{\epsilon\mu}$. That is
to say, strong chirality does not necessarily lead to strong
spatial dispersion.

Based on Eqs. (\ref{trans1})-(\ref{trans2}), we have computed
$\beta$ and $\epsilon_{DBF}/\epsilon$ or $\mu_{DBF}/\mu$ versus
$\kappa/\sqrt{\epsilon\mu}$, as shown in Figs. 1 and 2. When
$\kappa$ is very close to $\sqrt{\mu\epsilon}$, the value of
$\beta$ is quite large, indicating a strong spatial dispersion.
Hence the singular point is the very point of traditional
limitation. However, with $\kappa$ continuously increasing, the
spatial dispersion strength falls down very quickly. Therefore, if
$\kappa$ is not around $\sqrt{\mu\epsilon}$, e.g.
$\kappa<0.7\sqrt{\mu\epsilon}$ or $\kappa>1.3\sqrt{\mu\epsilon}$,
we need not take nonlinear terms into consideration at all. Hence
the strong spatial dispersion and nonlinearity cannot put the
upper limitation to chirality parameters either.

\begin{figure}[h,t,b]
\centerline{
\includegraphics[width=9cm]{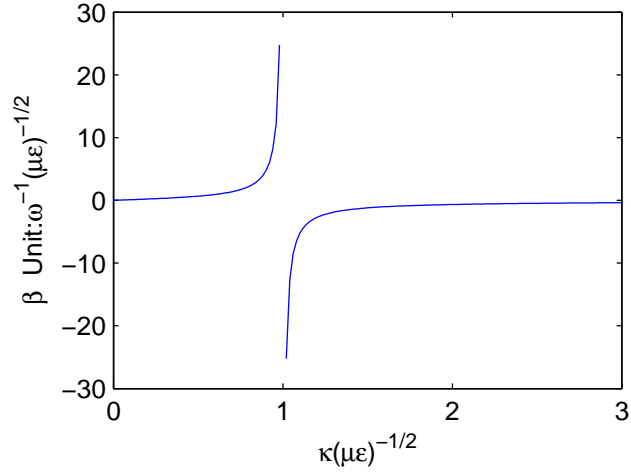}}
\caption{\small {The strength relationship of chirality and
spatial dispersion. The point of $\kappa/\sqrt{\mu\epsilon}=1$ is
singularity, corresponding infinite spatial dispersion coefficient
$\beta$. When $\kappa/\sqrt{\mu\epsilon}>1$, $\beta$ becomes
negative for keeping the positive rotation term coefficients with
negative $\epsilon_{DBF}$ and $\mu_{DBF}$.} \label{fig1}}
\end{figure}

\begin{figure}[h,t,b]
\centerline{
\includegraphics[width=9cm]{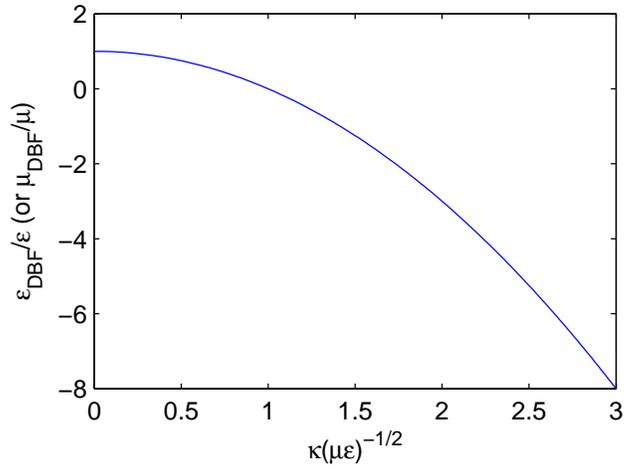}}
\caption{\small {With chirality strength increases,
$\epsilon_{DBF}$ and $\mu_{DBF}$ reduces quickly from $\epsilon$
and $\mu$ to $-\infty$.} \label{fig2}}
\end{figure}

When $\kappa$ is close to $\sqrt{\mu\epsilon}$ where the spatial
dispersion is strong, we need to take higher-order terms in the
DBF relations
\begin{eqnarray}
\bar{D} \hskip-2.5mm &=& \hskip-2.5mm \epsilon_{DBF}(\bar{E}+\beta_1\nabla\times\bar{E}+\beta_2\nabla\times\nabla\times\bar{E}+\dots),\\
\bar{B} \hskip-2.5mm &=& \hskip-2.5mm
\mu_{DBF}(\bar{H}+\beta_1\nabla\times\bar{H}+\beta_2\nabla\times\nabla\times\bar{H}+\dots),
\end{eqnarray}
where $\beta_{n}$ stands for the spatial dispersion of the $n$th
order. We remark that the above is different from the classical
nonlinear optics because it is strong spatial dispersion instead
of strong field intensity. Hence it is not a power series of
$\bar{E}$ and $\bar{H}$ fields.

Nevertheless, the Pasteur relations should remain the form as Eqs.
(\ref{cons1}) and (\ref{cons2}) as long as the medium is lossless
and reciprocal, no matter how strong the spatial dispersion is.
The only thing to be changed is the transform relation between DBF
and Pasteur models, which becomes much more complicated. That is
to say, though there are a lot of higher-order rotation terms,
$\bar{D}$ can still be represented as a real part proportional to
$\bar{E}$ and an imaginary part proportional to $\bar{H}$ with
modified coefficients. $\bar{B}$ has similar representations to
$\bar{D}$. The nonlinear terms contribute to the alteration of
effective $\epsilon$, $\mu$ and $\kappa$ in the Pasteur model,
which might be negative, leading to the energy problem again.

Actually, when introducing higher order terms in the DBF model,
$\epsilon_{DBF}$ and $\mu_{DBF}$ will be altered.Every rotation
term includes real and imaginary components, related to $\bar{E}$
and $\bar{H}$, respectively. Comparing to the DBF model, the
Pasteur model is relatively stable since its $\epsilon$ stands for
the total proportion of $\bar{D}$ to $\bar{E}$. Similar
conclusions are valid for $\mu$.

Moreover, it has already been shown that any medium satisfying the
Lorentz frequency-dispersive model has positive energy
densities.\cite{cui} Using the Pasteur relations, we have
\begin{eqnarray}
\frac{\mathrm{d}w}{\mathrm{d}t} \hskip-2.5mm &=& \hskip-2.5mm
\frac{\partial \bar{D}}{\partial t}\cdot \bar{E}+\frac{\partial
\bar{B}}{\partial t}\cdot
\bar{H}\nonumber\\
 \hskip-2.5mm &=& \hskip-2.5mm \frac{\mathrm{d} w'}{\mathrm{d}
t}+\mathrm{Re}\left(i\frac{\partial(\kappa\bar{H})}{\partial
t}\cdot\bar{E}-i\frac{\partial(\kappa\bar{E})}{\partial
t}\cdot\bar{H}\right)\nonumber\\
 \hskip-2.5mm &=& \hskip-2.5mm \frac{\mathrm{d} w'}{\mathrm{d}
t}+\kappa\mathrm{Im}\left(\frac{\partial\bar{E}}{\partial
t}\cdot\bar{H}-\frac{\partial\bar{H}}{\partial
t}\cdot\bar{E}\right),\label{energy}
\end{eqnarray}
in which $w'$ is the energy density in non-chiral Lorentz medium.
Substituting the relation $\bar{E}=\pm i\eta\bar{H}$ for two
circularly polarized eigenwaves into above equations, the last
term of Eq. (\ref{energy}) can be cancelled. Hence the energy
density remains the same as that in the common Lorentz medium.

\section{\bf Conclusions}

From Fig. 1, it is clear that enhancing spatial dispersion will
not lead to strong chirality and will reach the traditional
limitation point. This is why we have never succeeded in realizing
strong chirality no matter how to improve the asymmetry and
spatial dispersion.

Fortunately, as pointed out earlier, the strong chirality does not
require strong spatial dispersion. Hence the most important
difference between strong and weak chirality is that $\kappa$ and
$\beta$ have opposite signs, which necessarily leads to negative
$\epsilon_{DBF}$ and $\mu_{DBF}$. Here, $\kappa$ stands for
chirality and $\beta$ is the coefficient of the first order for
spatial dispersion. Strong chirality roots from using one type of
spatial dispersion to get the conjugate stereoisomer, or
chirality. It is an essential condition for supporting the
backward eigenwave in strong chiral medium.

In conclusion, a strong chiral medium behaves like Veselago's
medium. Under the weak spatial dispersion, the energy is always
positive for chiral medium. We show that strong chirality does not
equal strong spatial dispersion, which occurs only around a
singular point. Even in this small region with very strong spatial
dispersion, the Pasteur model is meaningful. Neither spatial
dispersion nor energy will hinder chirality to be stronger, but we
cannot realize strong chirality only by increasing the spatial
dispersion. The necessary condition of strong chiral medium is
that the chirality and spatial dispersion are of conjugated types.

We remark that strong chiral media have found wide applications in
the negative refraction and supporting of backward waves, which
have been discussed in details in Refs. \cite{sailing} and
\cite{4}-\cite{10}.

\section{\bf Acknowledgement}

This work was supported in part by the National Basic Research
Program (973) of China under Grant No. 2004CB719802, in part by
the National Science Foundation of China for Distinguished Young
Scholars under Grant No. 60225001, and in part by the National
Doctoral Foundation of China under Grant No. 20040286010.

\end{document}